\documentclass[12pt]{iopart}

\usepackage{iopams}
\usepackage{graphicx}
\usepackage{setstack}
\usepackage{color}

\begin{document}

\title{Correlated domains in spin glasses}

\author{Alain Billoire$^1$, Andrea Maiorano$^{2,3}$ and Enzo Marinari$^{2,4}$}
\address{$^1$ Institut de physique th\'eorique, CEA Saclay and CNRS, 91191
  Gif-sur-Yvette, France} 
\address{$^2$ Dipartimento di Fisica,  Sapienza Universit\`a di Roma, P. A. 
  Moro 2, 00185 Roma, Italy} 
\address{$^3$ Instituto de Biocomputaci\'on y F\'{\i}sica de Sistemas Complejos (BIFI)
  50018 Zaragoza, Spain} 
\address{$^4$ IPCF-CNR and INFN,
  Sapienza Universit\`a di Roma, P. A. Moro 2, 00185 Roma, Italy} 
\eads{alain.billoire@cea.fr, andrea.maiorano@uniroma1.it, enzo.marinari@uniroma1.it}

\begin{abstract}
We study the $3D$ Edwards-Anderson spin glasses, by analyzing
spin-spin correlation functions in thermalized spin configurations at
low $T$ on lattices of sizes up to $32^3$. 
We consider individual disorder samples and
analyze connected clusters of very correlated sites: we analyze how
the volume and the surface of these clusters increases with the
lattice size. We qualify the important excitations of the system
by checking how large they are, and we define a correlation length by
measuring their gyration radius. We find that the clusters have a very
dense interface, compatible with being space filling.
\end{abstract}
\maketitle 

\section{Introduction} 

The understanding of the low temperature behavior of spin glass
systems~\cite{BOOKS} is both a very interesting and challenging
problem and one of the most relevant open questions in classical
statistical mechanics. The mean field theory of the model has been
solved~\cite{PARISI}, and the rigorous mathematical proof that this
solution is correct~\cite{MATH} made the situation very clear.  In
finite dimensional spin glass however, the situation is completely
open, and there is no general agreement on what happens in the
infinite volume limit ($V\to\infty$). The two main possibilities are
on one side that in a finite dimension the model behaves as in mean
field~\cite{PARISI} (this is the replica symmetry breaking, RSB,
picture), by preserving all or most of the very peculiar features of
the mean field theory, and on the other side that the so called
droplet picture~\cite{DROPLET} (where only two stable states are
relevant for the critical behavior of the system) is realized.  For
some detailed analysis of the two points of view see for
example~\cite{REV2000,MAM}.  
The TNT scenario~\cite{MAM} suggests an intermediate situation (the
trivial-non-trivial picture) where the link overlap does not have a
complex behavior; here we only estimate the usual overlap. 
Even if it is very plausible that the
final answer to this crucial question will not come from numerical
experiments, it is clear that today large scale, large volume, very
accurate numerical simulations can allow to at least approach the
relevant low $T$, large $V$, regime, and give very relevant hints
about the physical behavior of such systems in the thermodynamic
limit. The numerical simulations of the Janus
supercomputer~\cite{janus:10,janus:10b,bfmmmy_2011,JANUS_2012} (a
special purpose computer~\cite{JANUS_HARD} specially built to simulate
very effectively spin systems with discrete variables and couplings)
have produced a large set of well thermalized spin glass
configurations on large (since we are considering spin glasses: the
same lattice sizes would not be so large for, say, an Ising model with
random field) systems at low $T$: our analysis will be based
on these spin configurations.

We take the approach, that has been discussed
in~\cite{bklm}, of considering local magnetizations and
correlation functions for a given disorder sample (without averaging over the disorder). This approach can give a very complete picture, and allow detecting
important features of the system. We will mainly use spin-spin
correlation functions, $k_{i,j}^{(J)}\equiv \langle\sigma_i
\sigma_j\rangle$ (where the brackets  indicate the thermal
average,  $i$ and $j$ are two lattice sites,  and  $(J)$ 
denotes a given realization of the couplings) and build upon how
much two sites are correlated in a given sample.

Using  the values of the $k_{i,j}$ we will build connected
clusters of correlated spins.  These clusters are related to
excitations that during the dynamics tend to move in a coherent way:
their spins are very correlated. In this sense these connected \
clusters are good proxies of \emph{droplets}
and studying their properties allows us to get information about the \emph{droplets}. In
ref.~\cite{KHMMP_2001} a similar approach was taken to study
excitations at $T=0$. It is thanks to the availability of well-thermalized
configurations from Janus for many disorder realizations that we are able to
analyze the finite small $T$ region.

\section{Model and data analysis} 

We study here the $3D$ Edwards Anderson spin glass with binary
couplings. We analyze well thermalized, equilibrium spin
configurations obtained with large scale numerical simulations by the
Janus collaboration~\cite{janus:10,janus:10b} based on parallel
tempering~\cite{PT} (see ref.~\cite{janus:10,janus:10b} for all the
parameters of the numerical simulation). We deal with $O(1000)$
samples of quenched disorder and many independent spin configurations
per disorder sample (for each lattice size), on simple cubic lattices
of sizes going up to $V=L^3=32^3$, for temperatures down to
$0.64\;T_c$, with periodic boundary conditions.

For a given disorder sample and lattice site $i$, we  form
the (connected) cluster ${\cal C}_i$ by adding recursively to ${\cal C}_i$ all
sites $j$ whose correlation function $\langle \sigma_j
\sigma_i\rangle$ is larger in absolute value than a given threshold
${\cal T}_{{min}}$. When this is done, we only keep the connected
component containing the site $i$.  The value of the threshold ${\cal
  T}_{{min}}$ is an important parameter of our construction: if it is
very small the cluster will fill all the available space, and it will
have no boundary, while if it is very large ${\cal C}_i$ will be small
(but its complement, formed from all lattice sites that do not belong
to ${\cal C}_i$, will be large). We would expect that in the
physically relevant range of values of ${\cal T}_{{min}}$ the
universal results of our computation (like, for example, the critical
exponents) should not depend on ${\cal T}_{{min}}$, and we will verify
that this is indeed the case.

For a given sample we will consider both the set of the $V$
(connected) clusters ${\cal C}_i$, analyzing their volume $N_i$
and their surface $S_i$ (defined as the size of the set of sites which
belong to the cluster and have at least one first neighboring site
that does not belong to it), and properties obtained by averaging over
the different clusters ${\cal C}_i$ (e.g. the cluster average volume
$N$ and average surface $S$). All these quantities should be labeled
also with an index $(J)$, denoting the disorder sample we are
considering, but we will omit this label for the sake of a clearer
notation. We will try to determine how quantities like $N$ and $S$ scale with
the lattice size. 

\section{Size and surface of connected clusters} 

\begin{figure}
\begin{center}
\includegraphics[width=0.5\textwidth,angle=0]{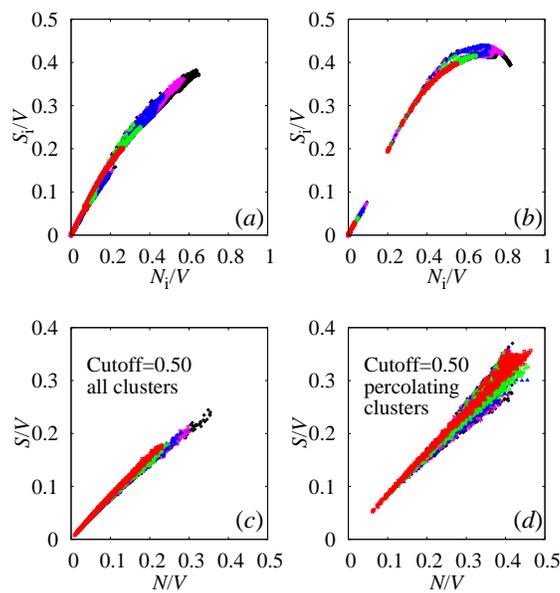}
\caption{In figures \ref{FIG_1}.(a) and \ref{FIG_1}.(b) we plot the
  quantities $S_i/V$ versus $N_i/V$ for all clusters ${\cal C}_i$
  ($i=1,\;...,\;V$) for two representative disorder samples. Symbols
  of different colors and shapes are used (like in figures
  \ref{FIG_1}.(c) and \ref{FIG_1}.(d)), for different values of the
  threshold ${\cal T}_{{min}}$: black 
  squares are for a threshold of $0.5$, cyan
  triangles pointing down $0.6$, blue triangles pointing up $0.7$,
  green circles $0.8$ and red squares $0.9$. In figure \ref{FIG_1}.(c)
  we show averages over all clusters ${\cal C}_i$ for a given sample
  (each point is for a given disorder realization, all disorder
  realizations are plotted), under the constraint $\lambda=0.5$ (see
  text): in figure \ref{FIG_1}.(d) we restrict the average to
  percolating clusters ($\lambda=0.5$).
\label{FIG_1}}
\end{center}
\end{figure}

We start by showing what happens in two representative disorder
samples in figure \ref{FIG_1}.a and \ref{FIG_1}.b. Both figures are
scatter plot that shows the cluster surface density $s_i\equiv S_i/V$
versus the cluster volume density $v_i\equiv N_i/V$ for cluster ${\cal
  C}_i$. Each point is for one of the $V$ clusters one can construct
when considering the disorder realization $J$. We show data for our
largest lattice, with $L=32$ (also, here and in the rest of this note,
$T=0.7\simeq 0.64 T_c$). Different colors are for clusters
reconstructed by using different values of ${\cal T}_{{min}}$. Scaling
for the different values of ${\cal T}_{{min}}$ is good: data points
with different colors fall on the same curve. The two samples we show
behave quite differently: in the one on the right (where larger
clusters are present: the maximum $n_i$ value is larger than $0.8$)
$s_i$ versus $n_i$ saturates and starts to decrease for $n_i\sim
0.7$. This decrease has a clear reason, due to finite size effects:
when clusters reach a size comparable to the lattice size (that has
periodic boundary conditions) it fills up all the available space, and it
has no space to have a boundary. As $n_i\longrightarrow 1$ the maximum
allowed surface density goes to zero. Because of this, we have
introduced a cutoff $\lambda$, and we discard from our analysis the
clusters such that $n_i>\lambda$. We have also analyzed the data for
different values of $\lambda$ in the range $(0.25,1.0)$, and our
claims will turn out not to depend sensitively on the choice of
$\lambda$. In the two lower figures \ref{FIG_1}.c and \ref{FIG_1}.d we
show averages of the cluster volume and surface density (with cutoff
$\lambda=0.5$) over all sites in individual realizations of the
disorder.  Each point represents one disorder sample. In the plot on
the left, an average is done over all clusters (with $\lambda=0.5$)
while on the  right the average is done over percolating clusters only
(defined as connected clusters that span the whole lattice in at least
one direction). When one considers all clusters, the small clusters,
containing a number of spin of order one, play an important
role. Selecting and analyzing instead only percolating clusters one is
keeping only the largest clusters (again with $\lambda=0.5$): in a
non-critical situations, with a finite correlation length and
correlations that fade away exponentially no such clusters would
exists in the infinite volume limit. The fact that we do find that for
increasing $V$ a finite fraction of clusters is indeed percolating is
a first signature of the fact that the basic, low energies domains are
highly non trivial sets of sites.  When we look at $s\equiv S/V$
versus $n\equiv N/V$ averaged over all clusters, we see that there is a
clear curvature: this is reasonable since for small clusters $S$ is
necessarily very similar to $V$. In the case of percolating clusters
one sees far less of a curvature: clearly here small clusters are
absent, since there is a minimal value of $N$ needed for percolating.

In order to proceed to a more quantitative assessment of the situation
we need to go a step further. Asymptotically one expects that for
large volumes $\overline{N}\propto L^{d_N}$, where the over-line is for
the average over the quenched disorder and $d_N$ is the exponent that
characterizes the asymptotic growth. We are ignoring here sub-leading
corrections since our numerical data would not allow to fit them (we
work with $L=16$, $24$ and $32$, i.e. we use three data points in each
fit): also it turns out that the best fits to the simple form with the
leading scaling behavior are very good, with low values of the
$\chi^2$. Analogously we define the exponent $d_S$ by the rate of
increase of the cluster surfaces with the volume, i.e.
$\overline{S}\propto L^{d_S}$. 

\begin{figure}
\begin{center}
\includegraphics[width=0.5\textwidth,angle=0]{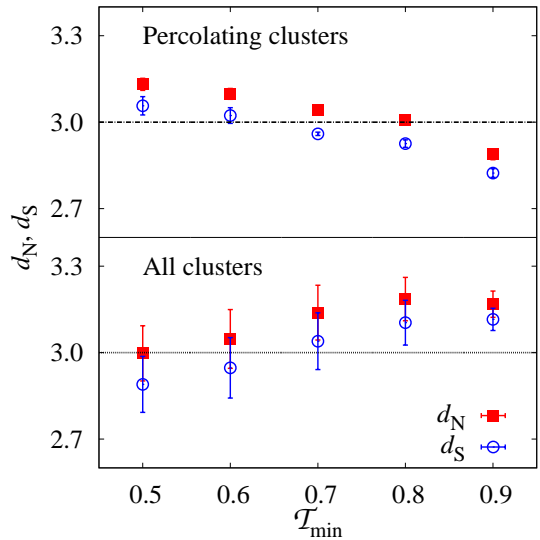}
\caption{The exponents $d_S$ and $d_N$ as a function of the
  correlation threshold ${\cal T}_{{min}}$, 
  considering either all connected clusters (bottom) or 
  percolating clusters only (top).\label{FIG_2}}
\end{center}
\end{figure}

By fitting our numerical data we obtain the values of the exponents
that we show in figure \ref{FIG_2}.  
In the case of the ferromagnetic Ising model our procedure, as applied
to the magnetization-magnetization connected correlation functions,
would find, for $T\ne T_c$, finite clusters of typical size $\xi$.
We plot the values of the
exponents in the two cases where we either consider all clusters, or
only percolating clusters. A few comments are in order. The dependence
of the exponents on the value of ${\cal T}_{{min}}$ is very weak: they
both stay in the range $(2.8,3.3)$ for values of ${\cal T}_{{min}}$
ranging from $0.5$ to $0.9$. In all these cases $d_N$ and $d_S$ are
very similar: their difference never exceed $0.1$. So, to summarize,
both exponents turn out to be stable as a function of ${\cal
  T}_{{min}}$ and $\lambda$, close to the values $3$ and very close
together. The low energy droplets in our system, at a value of the
temperature deep in the broken phase, appear to have, as far as we can
observe on the finite volume we can thermalize, a very diffused
interface, compatible with being space filling (namely $d_S=3$).

\section{Sponges} 

In a situation where the system is non critical we expect connected
clusters to have a (finite) size of the order of the correlation
length: survival, in the infinite volume limit, of extended clusters
is a signal of criticality. Such extended clusters can be defined, on
a finite lattice, in many different ways: we will use here a very
restrictive definition that we will denote by the word
``sponge''\cite{KHMMP_2001}. We declare a connected cluster to be
sponge-like if both it and at least one connected component of its
complement span the whole lattice in the $x$, $y$ and $z$
directions.  Only percolating clusters contribute to sponges.
We use the same cutoff $\lambda=0.5$, excluding clusters
that take a too large fraction of the lattice. This applies to ${\cal
  C}_i$ and to clusters in the complement of ${\cal C}_i$.

\begin{figure}[t]
\begin{center}
\includegraphics[width=0.5\textwidth,angle=0]{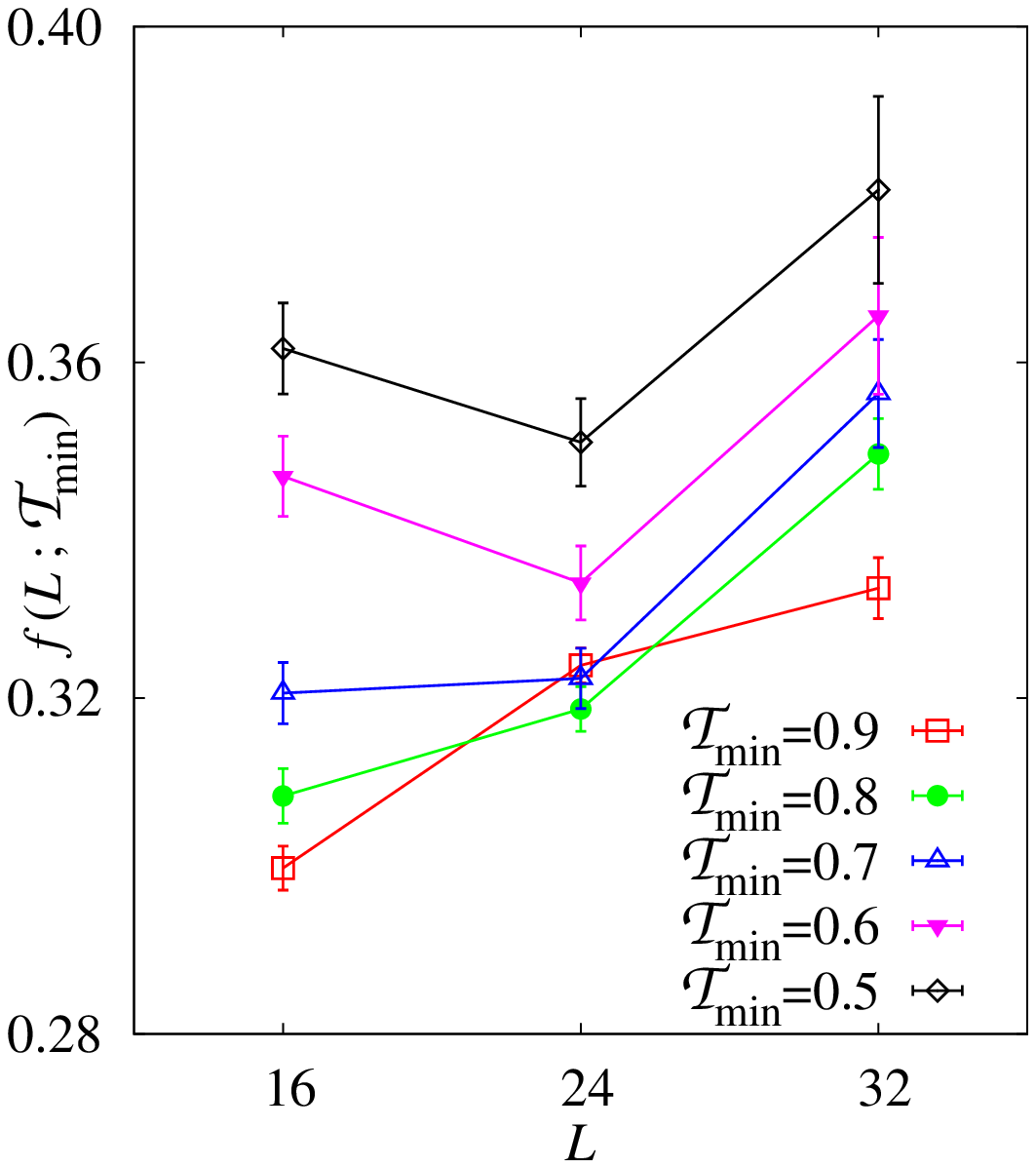}
\caption{\label{FIG_3}}
Fraction of sponges as a function of the (linear) lattice size.
\end{center}
\end{figure}

We show in figure \ref{FIG_3} the fraction of sponges as a function of
the (linear) lattice size. We use different values of 
${\cal T}_{{min}}$.  We compute the fraction of
sponges for a given disorder sample and we average this
fraction over the random quenched disorder.  The fraction of sponges
is sizable (of order $1/3$ with the very low cutoff $\lambda=0.5$) and
is never decreasing for increasing lattice size: the clear signal one
gets from these numerical data is that sponges will survive in the
infinite volume limit, and the typical configurations of the system
will be critical. It is clear that a numerical study of this
kind cannot, for many reasons, give a clear cut answer, but in the
very large volume range where the Janus numerical simulations allow us
to study equilibrium at low $T$ we are getting a clear hint towards
survival of extended excitations (using different values of $\lambda$
would not change the picture): this is what one would expect in a
situation where RSB holds. The same situation was found by analyzing
ground states of the system in ref. \cite{KHMMP_2001}. Here we are
working at finite $T<T_c$, with thermalized spin configuration. The
two results are in some sense complementary: we are interested in a
$T=0$ fixed point, so  we would like to be at low $T<T_c$, as close to
$T=0$ as possible (in principle at $T$ infinitesimally close to zero).
The analysis at $T=0$ gives very interesting information, but cannot
exclude the possibility that $T=0$ is a point with a special behavior.

\section{Typical lengths of non-extended clusters} 

\begin{figure}[t]
\begin{center}
\includegraphics[width=0.5\textwidth,angle=0]{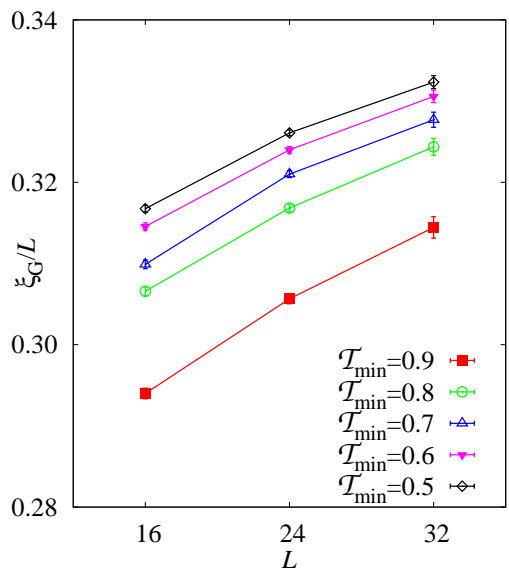}
\caption{\label{FIG_4}
$\xi_G/L$ versus $L$. The correlation length $\xi_G$ is defined 
from the non percolating cluster with the largest radius.}
\end{center}
\end{figure}

Last we investigate the geometrical structure of the connected
clusters, trying to determine a typical correlation length: we
summarize our results in figure \ref{FIG_4}. For each given disorder
samples we only consider
the clusters that do not percolate in any direction, and select the
one with the largest gyration radius:
$$
\xi_G^{(J)} \equiv \sqrt{\frac1N\sum_{k=1}^{N}r_k^{(2)}}\;,
$$ 
where $r_k^{(2)}$ is the square distance of site $k$ from the
center of mass of the cluster, and the sum runs over all sites of the
cluster. For fixed $T$ and $V$ 
we average $\xi_G^{(J)}$ over all disorder samples 
to obtain $\xi_G$, that we plot,
divided by $L$ in figure \ref{FIG_4}. We plot the results for
different values of ${\cal T}_{{min}}$ with
different symbols.

Selecting the largest, non percolating cluster, allows us to pick up
the largest physical length scale on a given finite lattice: the
procedure is analogous to the one used in a normal ferromagnet to
measure a physical, finite, correlation length below the critical
point.

In a spin glass we have a critical temperature and all temperature
values that are below it are possibly critical (i.e. they can have an
infinite correlations length): because of this the analysis is more
delicate than in a ferromagnet, We note first that the result does not
depend much on ${\cal T}_{{min}}$ and $\lambda$: the emerging picture
is really very stable.  A constant value of $\xi_G/L$ would imply a
divergence of the correlation length typical of a critical point: we
observe indeed a slight increase of this ratio, that we attribute to
finite size effects. Again, all the signatures we are detecting call
for a picture where a diverging length is determining the behavior of
the system. Using the non percolating cluster with the largest number
of sites, instead of the one with the largest radius, does give the
same kind of behavior.

\section{Conclusions} 

We have analyzed large-scale excitations of the system, estimated
critical exponents, the probability of finding large excitations, and
the gyration radius as an estimator of a correlation length.  We have
been able to work on thermalized spin configurations at low $T$ on
large lattices (experiments on spin glasses~\cite{ORBACH} show that we
are working on length scales not far from the physically relevant
ones). 

The emerging picture is strongly suggestive of the presence of RSB
(and compatible with a TNT scenario): typical excitations are large
and their interface is very dense, compatible with being, in the
infinite volume limit, space filling. In order to have exponents
compatible with the exponents we have measured, droplets should be
very different from the excitations originally proposed in the
``droplet picture''

\ack
We thank the Janus collaboration for allowing us to analyze their spin
configurations.  We acknowledge interesting discussions with Imre
Kondor, Victor Martin-Mayor and Sergio Perez-Gaviro. This work was
supported by the IIT Seed Project DREAM, IIT-Sapienza NanoMedicine Lab
and ERC contract no. 247328.

\end{document}